\documentclass[superscriptaddress,reprint,aps,pra]{revtex4-1}
\usepackage{graphicx}
\usepackage{amsmath}
\usepackage{multirow}
\usepackage{makecell}
\usepackage[hidelinks]{hyperref}
\usepackage[table,xcdraw]{xcolor}
\hypersetup{
    colorlinks,
    linkcolor={blue!80!black},
    citecolor={blue!80!black},
    urlcolor={blue!80!black}
}

\newcommand{\ket}[1]{\mbox{$\left|#1\right\rangle$}}
\newcommand{\bra}[1]{\mbox{$\left\langle#1\right|$}}

\newcommand{\set}[1]{\{#1\}}
\newcommand{\outerp}[2]{\ket{#1}\bra{#2}}

\newcolumntype{?}{!{\vrule width 2pt}}

\begin{document}

\title{Optimization and experimental realization of the quantum permutation algorithm}
\author{\.{I}. Yal\c{c}{\i}nkaya}
\affiliation{Department of Physics, Faculty of Nuclear Sciences and Physical Engineering, Czech Technical University in Prague, B\v{r}ehov\'a 7, 115 19 Praha 1-Star\'e M\v{e}sto, Czech Republic}
\email[]{iskender.yalcinkaya@fjfi.cvut.cz}
\author{Z. Gedik}
\affiliation{Faculty of Engineering and Natural Sciences, Sabanc{\i} University, Tuzla 34956, \.{I}stanbul, Turkey}

\date{\today}

\begin{abstract}

The quantum permutation algorithm provides computational speed-up over classical algorithms for determining the parity of a given cyclic permutation. For its $n$-qubit implementations, the number of required quantum gates scales quadratically with $n$ due to the quantum Fourier transforms included. We show here for the $n$-qubit case that the algorithm can be simplified so that it requires only $O(n)$ quantum gates, which theoretically reduces the complexity of the implementation. To test our results experimentally, we utilize IBM's $5$-qubit quantum processor to realize the algorithm by using the original and simplified recipes for the $2$-qubit case. It turns out that the latter results in a significantly higher success probability which allows us to verify the algorithm more precisely than the previous experimental realizations. We also verify the algorithm for the first time for the $3$-qubit case with a considerable success probability by taking the advantage of our simplified scheme.

\end{abstract}

\maketitle

\section{Introduction}

Quantum computers apparently will lead the way for information technology in the near future. Although it is not known whether they will completely replace classical computers, in theory, they already surpass their classical competitors in overcoming particular tasks such as integer factorization \cite{shor} and searching unsorted databases \cite{grover} as well as solving various optimization \cite{farhi, wiebe} and black box problems \cite{deutsch,bernstein,simon}. 

In mid 2016, IBM researchers brought a $5$-qubit quantum processor into service which is essentially composed of superconducting transmon qubits \cite{steffen,koch,ibmqx2,devoret,ibmqx4}. Anyone with no prior knowledge of the underlying hardware and/or experimental physics can program the processor via a graphical web interface allowing users to construct quantum circuits over a cloud server by simple mouse drags and drops. The circuits are sent to the server and queued to be  executed on the processor. The whole service, i.e., the processor and all of the other services provided for accessibility and documentation, is called IBM Quantum Experience (IQX) \cite{ibmqx}. It has already been used for conducting various research experiments including quantum teleportation \cite{fedortchenko,sisodia2017design},  Mermin \cite{alsina} and Leggett-Garg \cite{huffman} inequality violations, quantum cheque implementation \cite{behera}, quantum error correction, quantum arithmetic, quantum graph theory and fault-tolerant quantum computation \cite{devitt,vuillot}, Wigner function generation \cite{rundle}, nondestructive discrimination of Bell states \cite{sisodia2017experimental}, weight-four parity measurements \cite{takita}, and quantum uncertainty and measurement reversibility \cite{berta}. A comparison of the IQX and a trapped-ion based quantum computer has also been made by running a selection of quantum algorithms on both of them \cite{linke}. Most recently, IBM also included a $16$-qubit processor in IQX service which is available for beta access \cite{ibmqx3}.

The recently proposed quantum permutation algorithm (QPA) solves a black box problem two times faster than the best possible classical algorithm \cite{gedik}. It utilizes a single qudit to find out if the parity of a given cyclic permutation over $d$ elements is positive or negative. In a short time period, the algorithm has been verified by using various experimental methods, e.g., in $d=3$, by employing deuterium nuclei (spin-$1$) as NMR qutrits \cite{dogra}; in $d=4$, by employing sodium nuclei (spin $3/2$) as NMR ququarts \cite{gedik} as well as two-photon polarizations \cite{zhan}, single-photon polarization together with the spatial mode \cite{wang}, and orbital angular momentum of photons \cite{chen} as photonic ququarts.

In this article, we optimize the QPA by minimizing the number of required quantum gates when the algorithm is implemented by using an $n$-qubit register in $d=2^n$. To do that, we modify the original recipe by replacing the quantum Fourier transform (QFT) and its inverse with simpler transformations requiring quadratically fewer quantum gates. In parallel with this, we show that, instead of measuring all qubits at the end, it is enough to measure only one prespecified qubit for determining the parity of a given permutation. Since our optimized scheme has the potential of being beneficial for experimental setups consisting of $n$-partite qubit systems, we test it experimentally by using IQX. In this way, we verified the QPA in $d=4$ with a very high average success probability. We also examined the processor for its performance in realizing the original QPA in $d=4$ to make a comparison with the optimized case. Lastly, we realize the algorithm in $d=8$ and verify it with considerable success probability by using the optimized scheme.

This article is organized as follows: In Sec. \ref{sec:perAlg}, we briefly explain the QPA and introduce our optimized scheme. In Sec. \ref{sec:expReal}, we present our experimental results for $2$-qubit and $3$-qubit cases. In Sec. \ref{sec:conc}, we summarize and discuss our results.

\section{\label{sec:perAlg}Permutation problem}

Consider a family of linear transformations $\ket{\phi_m^\pm}=P_m^\pm\ket{\phi}$ with $m=0,1,\hdots,d-1$ in $d$-dimensional discrete space spanned by the orthonormal set of vectors $\mathcal{S}=\set{\ket{0},\ket{1},\hdots,\ket{d-1}}$ where $\ket{i}=(\delta_{0i},\delta_{1i},\hdots,\delta_{d-1,i})^T$. The operators $\hat{P}_m^\pm$ are $d\times d$ matrices and they permute the components of a given vector $\ket{\phi}$ in a cyclic manner as
\begin{equation}
P_m^\pm=\sum_{k=0}^{d-1}\outerp{{(m\pm k)}_{\text{mod}(d)}}{k}
\label{eq:permop}
\end{equation}
where $+$ and $-$ indicates the parity of the permutation. These cyclic permutations are indeed a subset of all possible permutations. More precisely, the components of $\ket{\phi_m^+}$ ($\ket{\phi_m^-}$) are obtained in ascending order by selecting adjacent components of $\ket{\phi}$ downwards (upwards) through the column vector in a periodic manner after starting with the $m$th component. One can interpret these permutations as being hidden in separate \textit{black boxes} which can only be viewed in terms of their inputs $\ket{\phi}$ and outputs $\ket{\phi_m^\pm}$. In this context, a single \textit{query} is defined as the examination of the output after the box is provided with an input. Thus, the permutation problem asks that, if we are given an arbitrary black box, how many queries are needed to estimate the parity of the corresponding permutation?

\subsection{Classical approach}

\begin{figure}
\centering
\includegraphics[scale=1.7]{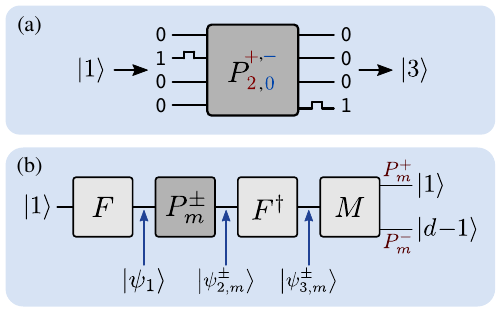}
\caption{(a) Demonstration of classical approach for solving the permutation problem in $d=4$. The input $\ket{1}$ gives the same output $\ket{3}$ for two different permutation operators $P_2^+$ and $P_0^-$, which points out the inadequacy of a single query for revealing the parity. (b) Circuit representation of the QPA for $d$ dimensions. After starting with the initial state $\ket{1}$, the permutation operator (black box) $P_m^{\pm}$ is sandwiched between quantum Fourier transform $F$ and its inverse $F^\dagger$. The measurement over the state $\ket{\psi_{3,m}^\pm}$ results in either $\ket{1}$ or $\ket{d-1}$ (up to some phase) with unit probability if the parity of the permutation is positive or negative, respectively.}
\label{fig:cpaqpa}
\end{figure}

A classical computer exploits a finite alphabet --- more specifically a binary alphabet $\set{0,1}$ --- for simulating any kind of computational algorithm.  Mathematically, this is equivalent to defining our vector space over the binary field, i.e., the components of any vector can be either $0$ or $1$. Furthermore, the vector components are interpreted as different ``wires'' entering the black box since they correspond to different physical states as shown in Fig. \ref{fig:cpaqpa}(a). It can be seen that a single query is not enough to determine the parity of the given permutation when we start with the basis state $\ket{1}$. Moreover, the same situation is valid for any other vector defined in this vector space.  Therefore, in the classical approach at least two queries with two different input vectors are required in order to solve the problem.

It is worth to emphasizing that, if we use another finite field with $d$ elements, we can solve the problem with a single query up to dimension $d$. This does not contradict the classical logic. However, the algorithm fails anyway in dimensions higher than $d$th. Enlarging the field where the vector space is defined does not provide a universal single-shot solution for the problem. Therefore, the binary field is chosen by convention.

\subsection{Quantum approach}

In the quantum approach, the vector space is defined over the complex field instead of the binary field as in the classical case. Therefore, the components of a vector correspond to the probability amplitudes which are represented by a single wire in the circuit diagram as shown in Fig. \ref{fig:cpaqpa}(b). The quantum algorithm starts to solve the problem by initializing the state 
\begin{equation}
\ket{\psi_1}=\frac{1}{\sqrt{d}}\sum_{k=0}^{d-1}{e^{i2\pi k/d}\ket{k}}.
\label{eq:psi1}
\end{equation}
Then, one of the permutation operators in Eq. \eqref{eq:permop} is applied which results in
\begin{equation}
\ket{\psi_{2,m}^\pm}=\frac{1}{\sqrt{d}}\sum_{k=0}^{d-1}{e^{i2\pi k/d}\ket{(m\pm k)_{\text{mod}(d)}}}.
\label{eq:psi2}
\end{equation}
Lastly, an inverse QFT is applied for decoding Eq. \ref{eq:psi2} into one of either
\begin{equation}
\ket{\psi_{3,m}^+}=e^{i2\pi m/d}\ket{1}
\label{eq:psi3+}
\end{equation}
or
\begin{equation}
\ket{\psi_{3,m}^-}=e^{i2\pi (d-1)m/d}\ket{d-1},
\label{eq:psi3-}
\end{equation}
if the parity of the given permutation is positive or negative, respectively. Therefore, a single query is adequate to solve the problem deterministically after measuring $\ket{\psi_{3,m}^\pm}$ in the $\mathcal{S}$ basis. The phases ($d$th roots of unity) uniquely assigned because the components of $\ket{\psi_1}$ play a key role in this computational speed-up by keeping track of the parity, which is classically impossible.

\subsection{Optimization of QPA for $n$-qubit cases}

\begin{figure}
\centering
\includegraphics[scale=1.1]{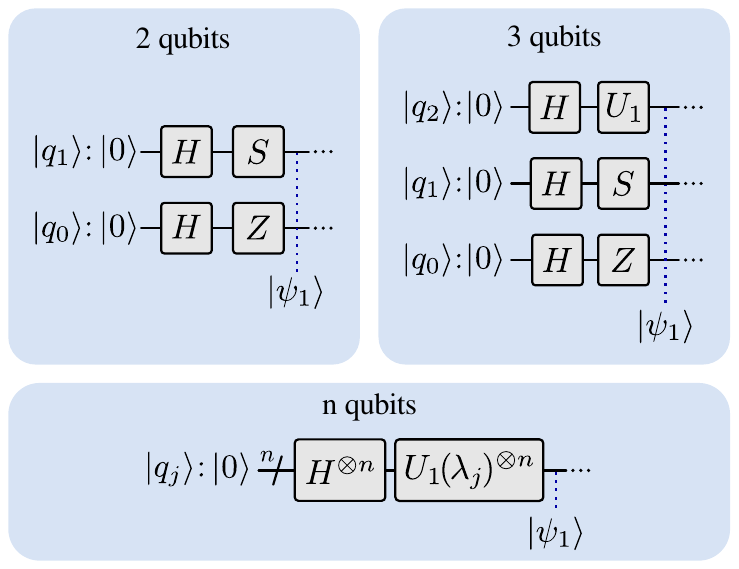}
\caption{Preparation of the $\ket{\psi_1}=\frac{1}{2}\left[\ket{0}+i\ket{1}-\ket{2}-i\ket{3}\right]$ state for $2$-, $3$-, and $n$-qubit cases. Here, $S$ and $Z$ are the phase and Pauli-$Z$ gates, which are special cases of $U_1$ given in Eq. \eqref{eq:hadps} with $\lambda=\pi/2$ and $\lambda=\pi$. The individual qubit states are denoted by $\ket{q_j}$, where $j=0,1,\hdots,n-1$.}
\label{fig:inist}
\end{figure}

The quantum Fourier transform for $n$ qubits is implemented by using $n(n+1)/2 $ Hadamard gates and controlled phase shift gates requiring QPA to include at least twice as much. Fortunately, a simplification can be done by completely getting rid of both the QFT and its inverse if the algorithm is implemented for $n$ qubits. In the following we interchangeably use notations $\ket{q}$ and $\ket{q_0q_1\hdots q_{n-1}}$ for $n$-qubit states, where $q=q_0 2^{n-1}+q_1 2^{n-2}+\hdots+q_{n-1} 2^0$. We first note that we always start by transforming the same initial state $\ket{1}\equiv \ket{00\hdots 01}$ to obtain the particular state $\ket{\psi_1}$ for any $d=2^n$, i.e., we do not actually need a general QFT capable to operate on all possible initial states. To see that let us write $F\ket{00\hdots 01}$ in the tensor product form of individual qubits as
\begin{equation}
\ket{00\hdots 01}\xrightarrow{F}\ket{\psi_1}=\frac{1}{\sqrt{2^n}}\bigotimes_{j=0}^{n-1}{\left\{\ket{0}+e^{i2\pi 2^{-(j+1)}}\ket{1}\right\}}.
\label{eq:qftof1}
\end{equation}
Thus, one can prepare the state $\ket{\psi_1}$ in Eq. \eqref{eq:psi1} by single qubit operations, i.e., first prepare all qubits in the ground state $\ket{0}$ and then, apply Hadamard and appropriate shift gates to each of them in a way to obtain $\ket{\psi_1}$ at the end. In other words, the effective transformation yielding $\ket{\psi_1}$ is 
\begin{equation}
\ket{\psi_1}=\left[\bigotimes_{j=1}^{n}HU_1(\lambda_j)\right]\overbrace{\ket{00\hdots 0}}^{\text{n}}.
\label{eq:iniStKron}
\end{equation}
Here, $H$ and $U_1$ are the Hadamard and phase shift gates which are given respectively as
\begin{equation}
H=\frac{1}{\sqrt{2}}
\begin{bmatrix}
1&1\\
1&-1
\end{bmatrix},~
U_1(\lambda_j)=
\begin{bmatrix}
1&0\\
0&e^{i\lambda_j}
\end{bmatrix},
\label{eq:hadps}
\end{equation}
with $\lambda_j=2\pi 2^{-j}$. Thus, the number of required gates for preparing the state $\ket{\psi_1}$ becomes $2n$ which is quadratically smaller than that of the original case. In Fig. \ref{fig:inist}, quantum circuits preparing $\ket{\psi_1}$ for $2$-, $3$-, and $n$-qubit cases are provided. We note for the $2$-qubit case that the original QFT scheme apparently requires one gate less than that of the optimized scheme. However, as we will see in Sec. \ref{ssec:2qoc} that the implementation of a controlled phase gate in the quantum processor is demanding in terms of the gates required. Therefore, in Sec. \ref{sec:expReal}, we will keep using the prescription given in Fig. \ref{fig:inist} for initializing the $2$-qubit case experimentally.

\begin{figure}
\centering
\includegraphics[scale=2]{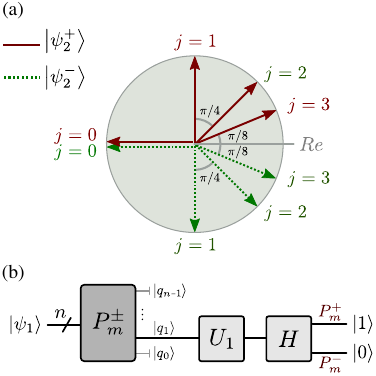}
\caption{Complex-plane representation of the phases $e^{\pm i\theta_j}$ in Eq. \eqref{eq:relPhInd} up to $4$ qubits. The conjugate phases correspond to different parities. Only the first qubit ($j=1$) can be found in one of the two orthogonal states $\ket{0}\pm i\ket{1}$ depending on the parity in any dimension $d$. (b) Circuit representation of the optimized QPA for $n$ qubits after the state preparation (see Fig. \ref{fig:inist}). Here, $U_1$ is the phase gate given in Eq. \eqref{eq:hadps} with $\lambda=\pi/2$. After the Hadamard gate, only $\ket{q_1}$ is measured and the other qubits are neglected. The measurement results in either $\ket{0}$ or $\ket{1}$ with unit probability if the permutation is negative or positive, respectively.}
\label{fig:relph}
\end{figure}

Further optimization can be done by eliminating the inverse QFT. We know that the algorithm deterministically results in either \eqref{eq:psi3+} or \eqref{eq:psi3-}. Therefore, before the inverse QFT,  $\ket{\psi_{2,m}^\pm}$  can be chosen as  $F\ket{1}\equiv\ket{\psi_{2,m}^+}$ and $F\ket{d-1}\equiv\ket{\psi_{2,m}^-}$ up to some phases depending on $m$, respectively. We neglect these global phases since they have no effect on measurements and we will remove the subindex $m$ to indicate this. The tensor product form of $\ket{\psi_2^+}=\ket{\psi_1}$ has already been given in Eq. \eqref{eq:qftof1}. Similarly, $\ket{\psi_2^-}$ can be written as
\begin{equation}
\ket{11\hdots 1}\xrightarrow{F}\ket{\psi_2^-}=\frac{1}{\sqrt{2^n}}\bigotimes_{j=0}^{n-1}{\left\{\ket{0}+e^{-i2\pi 2^{-(j+1)}}\ket{1}\right\}}.
\label{eq:qftof2}
\end{equation}
We see by comparing Eq. \eqref{eq:qftof1} and Eq. \eqref{eq:qftof2} that $\ket{\psi_2^+}$ and $\ket{\psi_2^-}$ differ from each other only by the relative phases possessed by the individual qubit states 
\begin{equation}
\ket{q_j^\pm}=\ket{0}+e^{\pm i\theta_j}\ket{1}.
\label{eq:relPhInd}
\end{equation}
Here, $\theta_j = 2\pi 2^{-(j+1)}$ and the phases $e^{\pm i\theta_j}$ are drawn in Fig. \ref{fig:relph}. It is seen that, independent of $n$, the first qubit is always found either one of the two states $\ket{q_1^\pm}=\ket{0}\pm i\ket{1}$ regarding to the parity. Thus, one can immediately solve the problem by performing a Pauli $Y$ measurement on the first qubit. However, sometimes it is experimentally more convenient to measure the qubits in the basis in which they have been prepared. For example, in the quantum processor we are only allowed to perform Pauli $Z$ measurements. Therefore, before the measurement in the computational basis $\set{\ket{0},\ket{1}}$, we sequentially apply $U_1(\pi/2)$ and $H$ operators given in Eq. \eqref{eq:hadps} as
\begin{equation}
\ket{\Psi_2^\pm}\xrightarrow{U_1}{\ket{0}\mp\ket{1}}\xrightarrow{H}
\begin{cases}
\ket{0}, & \text{for~} - \\
\ket{1}, & \text{for~} +
\end{cases}
\end{equation}
for eventually obtaining deterministic results (see Fig. \ref{fig:relph}(b)), which is analogous to the task of the inverse QFT in the original scheme. We obtain either $\ket{0}$ or $\ket{1}$ with certainty for negative and positive permutations, respectively, by a single query. We note that the other qubits with $j>1$ cannot be used for the same purpose since they do not possess orthogonal states for different parities.

\section{\label{sec:expReal}Experimental realization}

We performed $3$ different experiments by using the IQX web interface where the physical qubits in the processor are labeled by $\mathtt{q[i]}$ with $i=0,1,2,3,4$ and they are initially found in the ground state $\ket{0}^{\otimes 5}$. We initialize all experiments by preparing $\ket{\psi_1}$ according to Fig. \ref{fig:inist}, and then we complete the circuits as shown in Fig. \ref{fig:relph}(b) for the optimal cases (we provide further information about the original case in Sec. \ref{ssec:2qoc}). For each $P_m^\pm$, we build separate circuits which are realized $8192$ times by the processor with a runtime of one minute or less. Consequently, we obtain finite-sample (probability) distributions over the possible measurement outcomes with the sample size $8192$ and we average each distribution over $5$ samples ($5 \times 8192$ realizations in total). Since we run all circuits one after another, we assume that sequential realizations does not affect each other, i.e., the processor is prepared exactly in the same initial state for each circuit.

We note that qubits in the processor are slightly different from each other in terms of their calibration parameters such as energy relaxation time, coherence time, gate error and readout error. The coherence of the processor is gradually lost as the number of gates in the circuit is increased. Therefore, the results vary depending on which qubits have been used and how many gates have been implemented for each of them. For this reason, we chose the qubits providing the most efficient and the most robust results for the calibration parameters at the time we made the experiment. We provide these parameters together with a very brief technical description of the processor in the appendix.

\begin{figure}[b]
\centering
\includegraphics[scale=1]{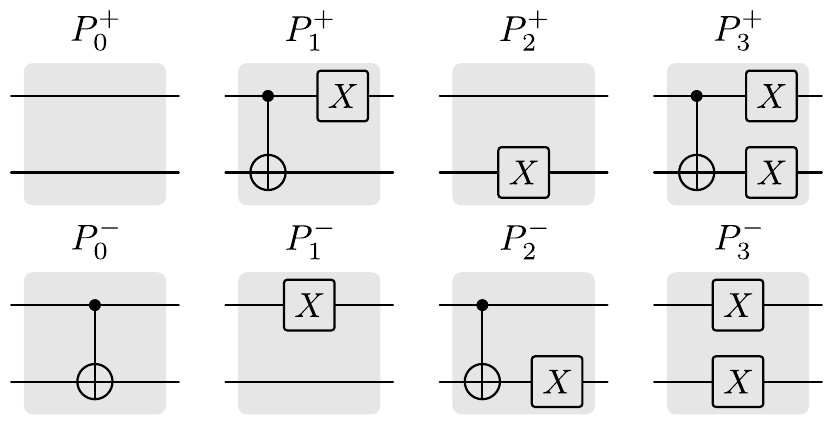}
\caption{Circuit diagrams of all $8$ permutation operators $P_m^\pm$ for $n=2$. Here, one-qubit and two-qubit gates are the Pauli $X$ and  the controlled NOT gates, respectively. The lower (upper) line corresponds to $\ket{q_0}$ ($\ket{q_1}$).}
\label{fig:2qpermop}
\end{figure}

\subsection{\label{ssec:2qc}Optimal QPA for the 2-qubit case}

\begin{figure*}
\centering
\includegraphics[scale=1.1]{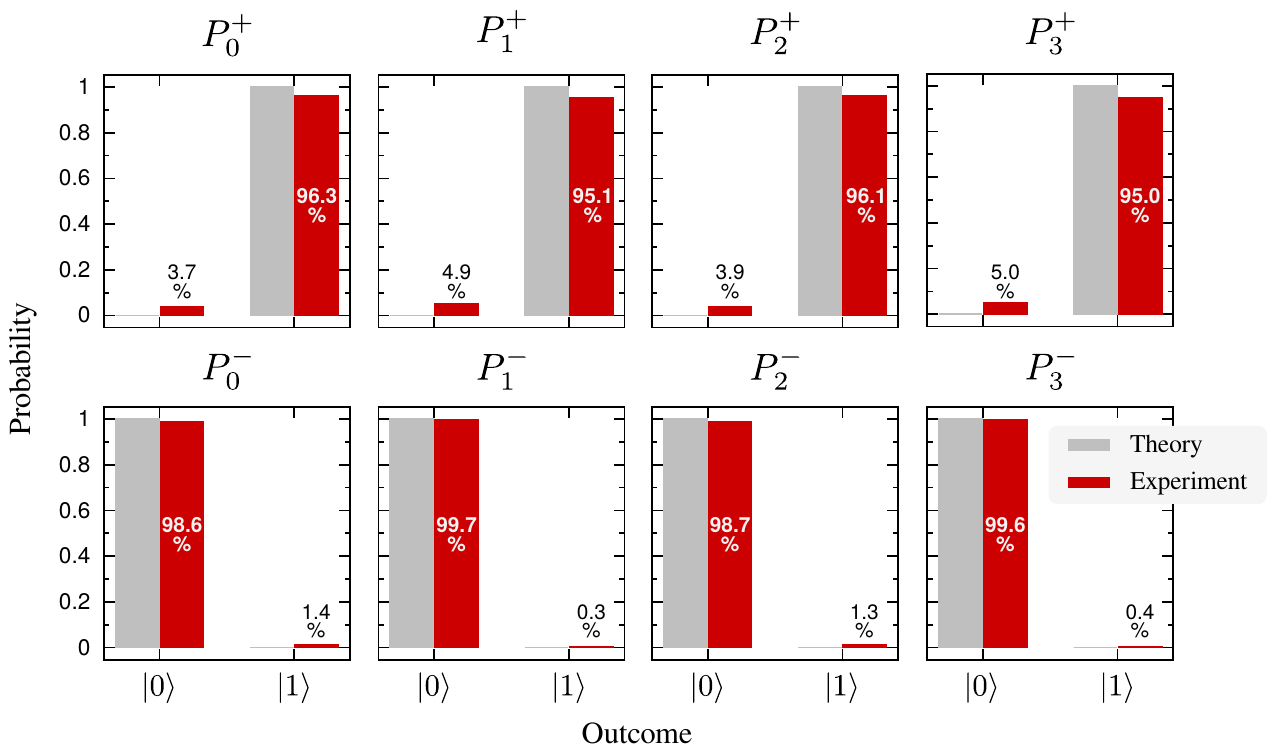}
\caption{Experimental results for $2$-qubit case. In agreement with the theory, the measurement yields $\ket{0}$ ($\ket{1}$) when the parity of a given permutation negative (positive) with an average success probability of $97.4\%$. Since the standard errors of individual sample means are at most $\pm 0.120\%$ for the sample size $8192$, we do not include the error bars.}
\label{fig:2qres}
\end{figure*}

We choose $\mathtt{q[1]}$ and $\mathtt{q[0]}$ as $\ket{q_0}$ and $\ket{q_1}$, respectively. This opposite labeling can be seen unnecessarily disorienting at the first glance. However, it is required for the consistency between our formalism and the architecture of the processor when the performance issues are considered. We implement $8$ permutation operators $P_m^\pm$ as shown in Fig. \ref{fig:2qpermop}. The Pauli $X$ gates appear on the $\ket{q_0}$ line may seem unnecessary since $\ket{q_1}$ is measured at the end. However, we do not eliminate these gates from circuits for further optimization because of the assumption that we are not allowed to ``look'' inside of the black box. Therefore, we do not intervene in the structure of the permutation operators and we use them as they are.

The probability distributions for each $P_m^\pm$ are shown in Fig. \ref{fig:2qres}. It is seen that we obtain high fidelity results verifying the theory, which are slightly different from each other.  Therefore, we can assign an individual success probability $S_m^\pm$ to each $P_m^\pm$, which is the chance of finding the parity with a single query, e.g., $S_0^+=0.963$ and $S_0^-=0.986$. The differences in success probabilities occur due to two possible reasons: i) The processor is less successful in one or more of preparing, protecting and measuring the excited state $\ket{1}$, which eventually causes a $3.5\%$ drop in the success probabilities of the positive parities on average (see the Appendix for the relaxation times of individual qubits). We verified this fact by by doing independent experiments where we examine the ground and excited states in detail. ii) The permutation operators induce different decoherence rates on the processor. Particularly, the existence of controlled NOT gates reduces the success probabilities about ${1\%}$ when we compare the results for a given parity. The reason for this is that controlled NOT gate has approximately $10$ times higher error rate than that of the one-qubit gates. 

In terms of computational issues, the exact parity determination with the processor for an unknown $P_m^\pm$ is pointless since one needs more than one realizations to obtain a reliable result, which consequently provide no speed up in practice. On the other hand, the individual success probabilities $S_m^\pm$ given in Fig. \ref{fig:2qres} do not provide a precise information for single-shot realizations since we do not know which permutation operator we have queried at all. If we define an average success probability $\bar{S}$ over all permutation operators instead, it would be meaningful in the case of an ensemble of different permutation operators rather than a single one. For these reasons, to provide computational insight to our results, let us consider the scenario as follows: We have a set of $N$ unknown permutation operators $\set{(P_m^j)_1,\hdots,(P_m^j)_N}$ whose elements are distributed according to a probability distribution $p_m^j$ with $j\in\set{+,-}$. Our aim is to find the parity of each operator via one-shot experimental realizations, which amounts to $N$ realizations in total. Now, we can define the average success probability as $\bar{S}=\sum_{m,j}{p_m^j S_m^j}$. If $p_m^j$ is uniform and $N\gg 1$ \footnote{Here, the condition $N\gg 1$ ensures that sample means are normally distributed. More precisely, we assert the condition $N>9dC/(1-C)$ with $C=\text{max}{(S_m^j)}$ so that the interval $1-\text{max}(S_m^j)$ becomes greater than three-sigma intervals of the sample means.}, we obtain $\bar{S}=0.974$ with the average standard error 
\begin{equation}
\bar{S}\pm\frac{1}{\sqrt{Nd}} \sum_{m,j}\left[S_m^j-(S_m^j)^2\right]^{1/2}.
\end{equation}
For $N=10^5$, for example, one can be sure that $97.4\%$ of $N$ results are correct with an error margin of $\pm 0.130\%$. Therefore, we are able to classify $N$ permutation operators according to their parities by $N$ realizations where the same process would require at least two times more operations in case of any classical algorithm. Nevertheless, approximately $2.6\%$ of accuracy is lost due to decoherence effects in the processor and measurement errors. In general, $P_m^{j}$'s are not necessarily distributed uniformly and we may have no idea about this distribution at all. Therefore, in a more realistic scenario, the mean $\bar{S}$ practically varies within the interval $\left[0.950,0.997\right]$ as long as $N\gg 1$, where the bounds are fixed by $P_3^+$ and $P_1^-$.

\subsection{\label{ssec:2qoc}Original QPA for the 2-qubit case}

\begin{figure*}
\centering
\includegraphics[scale=1.1]{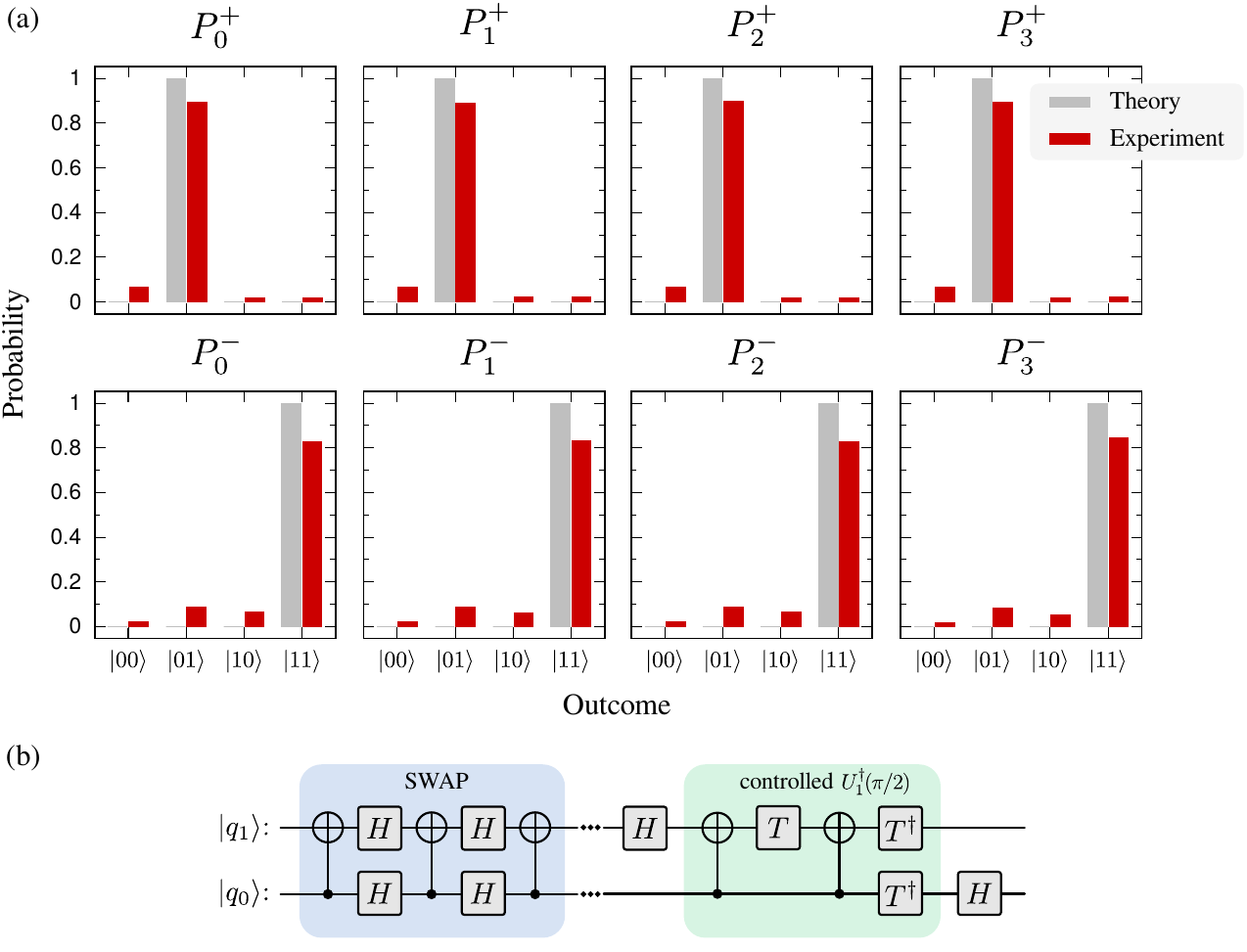}
\caption{(a) Experimental results for the original QPA for $2$-qubits. In agreement with the theory, the measurement most likely yields either $\ket{01}\equiv\ket{1}$ or $\ket{11}\equiv\ket{3}$ for positive and negative parities, respectively, with an average success probability of $86.3\%$. The error bars are not included since the standard errors of sample means are at most $\pm 0.349\%$ for the sample size $8192$. (b) An efficient circuit diagram of inverse QFT ($F^\dagger$) for the processor \cite{amy,nam}. The qubit $\ket{q_0}$ ($\ket{q_1}$) corresponds to $\mathtt{q[3]}$ ($\mathtt{q[2]}$) in the IQX web interface. Here, $H$ and $T$ are the Hadamard and $U_1(\pi/4)$ phase gates as given in Eq. \eqref{eq:hadps}, respectively. The SWAP gate is omitted in the actual implementation for efficiency. Instead, the lines are interchanged before the dots.}
\label{fig:2qorgres}
\end{figure*}

We choose $\mathtt{q[3]}$ and $\mathtt{q[2]}$ as $\ket{q_0}$ and $\ket{q_1}$, respectively. As already mentioned in Sec. \ref{sec:perAlg} that the original scheme requires an $F$ to initialize the known state $\ket{\psi_1}$ and an $F^\dagger$ to decode the unknown state $\ket{\psi_2^\pm}$ for the measurement (see Fig. \ref{fig:cpaqpa}). We keep using the procedure given in Fig. \ref{fig:inist} to prepare the state $\ket{\psi_1}$ because there is still no reason to use an $F$ to obtain a known state. However, an $F^\dagger$ is inevitable because of the fact that we need a general transformation acts on an unknown state. There is no physical gate or a predefined subroutine in the IQX interface for $F^\dagger$. Nevertheless, it can be efficiently implemented as shown in Fig. \ref{fig:2qorgres}(b) \cite{amy,nam}. We note here that the controlled NOT gates can only be placed in the $\ket{q_0}$ (control) to $\ket{q_1}$ (target) direction because of the architecture of the processor (see the coupling maps in the Appendix). For this reason, the middle controlled NOT included in the SWAP gate is inverted via four additional $H$ gates. Furthermore, if we implement the $F^\dagger$ exactly as shown, each permutation operator including a controlled NOT gate would also require these additional $H$ gates (see Fig. \ref{fig:2qpermop}). However, in our implementation of $F^\dagger$, we discard the SWAP gate and instead, we interchange the two lines $\ket{q_0}$ and $\ket{q_1}$ before the horizontal dots appear in Fig. \ref{fig:2qorgres}(b), i.e., both the state preparation and the permutation operator circuits are implemented upside down. Thus, we only use two controlled NOT and five one-qubit gates for $F^\dagger$ and we do not need extra $H$ gates for the permutation operators anymore. After $F^\dagger$, we measure both qubits in the computational basis separately to obtain results corresponding to the basis $\left\{ \ket{00},\ket{01},\ket{10},\ket{11} \right \}$. We note that we have to use more than $n(n+1)/2=3$ gates to implement $F^\dagger$ since the processor does not perform a controlled phase gate by default. 

In Fig. \ref{fig:2qorgres}(a), the average probability distributions for different $P_m^\pm$ are given. Although the results are pretty much consistent with the theory, the average success probability is reduced about $11\%$ in comparison with Fig. \ref{fig:2qres}. However, this is not surprising since we utilize two controlled NOT gates and five one-qubit gates for implementing the $F^\dagger$ and we also measure qubit $\ket{q_0}$ at the end. These operations cause additional errors which do not appear in the optimal case at all. Actually, we can estimate the error probability of $F^\dagger$ by using Fig. \ref{fig:app} in the Appendix as $4.1\%$. By also considering the difference between the measurement (readout) errors of the two cases, we end up with a total error probability of $9.2\%$ which approximately explains the decrease in the success probability. 

\subsection{\label{ssec:3qc}Optimal QPA for the 3-qubit case}

\begin{figure*}
\centering
\includegraphics[scale=1.1]{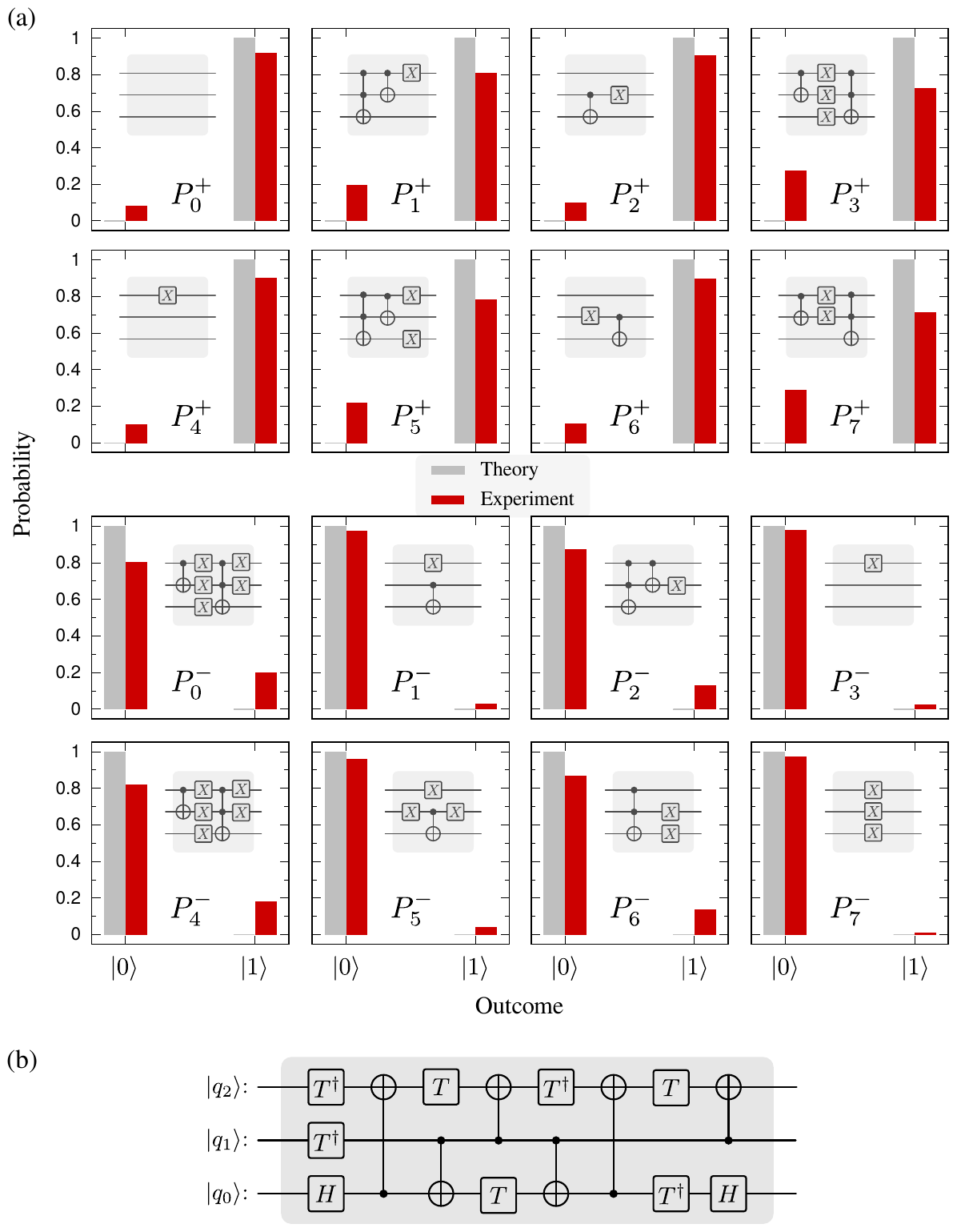}
\caption{(a) Experimental results for $3$-qubit case. In agreement with the theory, positive (negative) parities most likely end up with the state $\ket{1}$ ($\ket{0}$) with an average success probability of $86.8\%$. We do not include the error bars since the standard errors of the sample means are at most $\pm 0.500\%$ for the sample size $8192$. (b) Implementation of Toffoli gate \cite{amy} with the target on $\ket{q_0}$. Here, $H$ and $T$ are the Hadamard and $U_1(\pi/4)$ phase gates as given in Eq. \eqref{eq:hadps}, respectively.}
\label{fig:3qres}
\end{figure*}

We choose $\mathtt{q[4]}$, $\mathtt{q[3]}$ and $\mathtt{q[2]}$ as $\ket{q_0}$, $\ket{q_1}$ and $\ket{q_2}$, respectively. The circuit diagrams of $16$ permutation operators are given in Fig. \ref{fig:3qres}(a) together with the corresponding probability distributions. It is seen that in order to realize the half of permutation operators, we need Toffoli gates which are not physically implemented in the processor. For this reason, we use a combination of six controlled NOT and nine single-qubit gates to implement it as shown in Fig. \ref{fig:3qres}(b). By using Fig. \ref{fig:app} in the Appendix, we can calculate the error probability of the Toffoli gate as approximately $19\%$ which is significantly large. 

The probability distributions in Fig. \ref{fig:3qres}(a) significantly verify the theory with an average success probability of $\bar{S}=0.868$. From the computational point of view, after processing $N=10^5$ homogeneously distributed unknown permutation operators, one can be sure that $86.8\%$ of the results are correct with an error margin of $0.56\%$ (see Sec. \ref{ssec:2qc} for details). For a nonhomogeneous distribution instead, the mean $\bar{S}$ practically varies within the interval $[0.713,0.978]$ where the bounds are determined by $P_7^+$ and $P_3^-$. Compared to optimal $2$-qubit case, the average success probability is less and the average standard error is more due to the high error probability of the Toffoli gate. It can also be seen from Fig. \ref{fig:3qres}(a) that the individual success probabilities get considerably reduced for the cases where a Toffoli gate is used. Therefore, the results for $3$-qubit case can be considered too erroneous for claiming a quantum speed-up in practice. However, they still provide a solid proof for the verification of the QPA in $d=8$ while the original recipe might result in completely incorrect results due to the vast amount of quantum gates it requires.

\section{\label{sec:conc}Conclusion}

We introduced an optimization method for the QPA for $n$-qubit cases by showing that the required number of quantum gates for implementing the algorithm can be reduced so that it scales with $n$ instead of $n^2$ as in the original recipe. This result points out a quadratic decrease in the amount of the sources for an experimental setup. It is further not required to measure each individual qubit to find out the parity of the permutation, i.e., a single measurement on a particular qubit is adequate. By utilizing IBM's quantum processor, we have demonstrated the significant experimental advantage of the optimized scheme over the original scheme for the $2$-qubit case. Moreover, we tested and verified the QPA in $d=8$ for the first time with an appreciable success probability.

Our results indicate that the processor can practically be used approximately with $97\%$ efficiency to classify large number of unknown permutation operators according to their parities by exploiting two-to-one speed-up ratio compared with any classical algorithm. For the $3$-qubit case, this efficiency reduces to approximately $87\%$ mainly because of our costly implementation of the Toffoli gate. For this reason, a physically integrated Toffoli gate would be a significant improvement for the IBM's quantum processor for realizing any kind of reversible Boolean function similar to the permutation operator. 

In conclusion, as the quantum computers are being improved by reducing the error rates, it will be possible to implement more complex algorithms in larger dimensions. Based on our results, we believe that gate-model-based quantum computers are promising candidates for the future of quantum computation technology.

\appendix*

\section{\label{app:techdet}Technical Details about the IBM's 5-Qubit Quantum Processor}

The qubits in the processor are composed of superconducting inductors, capacitors, and Josephson junctions. The Josephson junctions make the system's potential anharmonic and give rise to energy levels that are not equally spaced. In this way, one can consider the system as a qubit with the unique zero to one transition frequency $f$. Two important parameters for the qubits are the relaxation time $T_1$ and the dephasing time $T_2$, where $T_1$ is the decay time of a qubit from its excited state $\ket{1}$ to the ground state $\ket{0}$ while $T_2$ quantifies the time it takes for a superposition state to lose its phase relationship. Here, $T_1$ can be one of the several reasons making the processor behave slightly state dependent as we discussed in Sec. \ref{ssec:2qc}. The qubits are coupled with each other via two coplanar waveguides making the implementation of the controlled NOT gates possible. The qubit with higher (lower) frequency becomes the control (target) qubit and the full control-to-target coupling maps are given by $\set{0:(1,2),1:(2),3:(2,4),4:(2)}$ and $\set{1:(0), 2:(0,1,4),3:(2,4)}$ for the two processors called ibmqx2 and ibmqx4, respectively. Both chips have the same connectivity except the directions in their coupling maps (the IQX team first introduced ibmqx2 and then replaced it with ibmqx4 in October, 2017). The restriction in the coupling directions mainly determined our way of implementing the permutation operators, the Toffoli gate, and the inverse QFT in the article. The error probabilities of controlled NOT gates are denoted by $e_g^{ij}$ where $i$ and $j$ represents the control and target qubits, respectively. Each qubit also has a dedicated coplanar waveguide for realizing the single-qubit operations and readout, where the respective error probabilities are denoted by $e_g$ and $e_r$. All systems are kept in a refrigerator with very low temperatures on the order of milli-Kelvins. Detailed information about the processor and the superconducting quantum computers can be found in Refs. \cite{ibmqx2,steffen, devoret, gambetta,ibmqx4}

\begin{figure}
\centering
\includegraphics[scale=1.03]{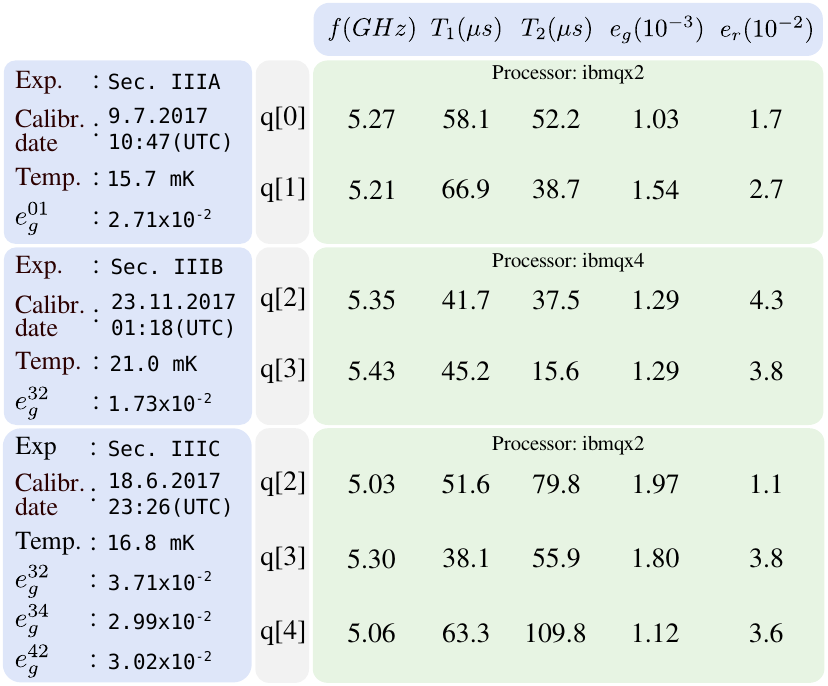}
\caption{Calibration parameters of the processors, namely ibmqx2 and ibmqx4, during we perform the experiments in Secs. \ref{ssec:2qc}, \ref{ssec:2qoc}, and \ref{ssec:3qc}. Here, $f$, $T_1$, $T_2$, $e_g$, and $e_r$ are the qubit frequency, relaxation time, coherence time, single-gate error and readout error, respectively. The controlled NOT gate errors are denoted by $e_g^{ij}$ where the target qubit is $j$.}
\label{fig:app}
\end{figure}

The IQX team recalibrate the processor at least once a day and the above-mentioned parameters change slightly. In Fig. \ref{fig:app}, we provide these parameters for three different days in which we performed our experiments. 

\begin{acknowledgments}
We would like to thank D. Maslov for pointing out the typos in the manuscript and for his helpful comments. We also acknowledge use of the IBM Q Experience for this work. The views expressed are those of the authors and do not reflect the official policy or position of IBM or the IBM Q Experience team. \.{I}.~Y.\ was partially supported by the project RVO 68407700.
\end{acknowledgments}

\bibliography{bibliography}

\end{document}